# William Maximilian Lindley: Fifth Director of the BAA Variable Star Section

Jeremy Shears

## Abstract

William Maximilian Lindley, MC, MA, FRAS, AMICE (1891-1972) served as fifth Director of the BAA Variable Star Section from 1939 to 1958. He was an active variable star observer for many years and he wrote numerous publications on the observations made by Section members. This paper discusses Lindley's life and work, with a particular focus on his contribution to variable star astronomy.

## Introduction

The British Astronomical Association's Variable Star Section (BAA VSS), launched in 1890, is the world's longest established organisation for the systematic observation of variable stars (1). William Maximilian Lindley (1891 – 1972; Figure 1), Max Lindley as he was usually called, became its fifth Director in 1939 and remained in office until 1958, making him the longest serving VSS Director to date. Prior to this he was VSS Secretary for several years. An engineer by profession, Lindley also served in the Army during both World Wars. He spent most of his life at Trevone, near Padstow, on the north Cornwall coast. Lindley's obituary, written by Gordon Patston (1902-1989) who knew Lindley for nearly 40 years largely through the VSS, was published in the Journal in 1973 (2).

## The Lindleys: a family of engineers

Max Lindley was born at Frankfurt-am-Main, Germany, on July 27 1891. He had two sisters, Julia and Ottelie. His father, Sir William Heerlein Lindley (1853-1917; Figure 2) was a well-known civil engineer who specialised in the design and construction of sanitation systems in cities across Europe (3). His projects encompassed the construction of water pipelines, sanitation systems and waterworks in some 35 European cities. These included the waterworks in Warsaw, the sewage works in Prague, which were built between 1895 and 1906 and were in use until 1967, and in 1909 he designed the water and sewerage system for Łódź, Poland, although the project wasn't implemented until the 1920s. However, W.H. Lindley's most ambitious project was the water supply system for Baku in Azerbaijan (4), which he worked on from 1899 to his death in 1917 (5). Baku was the centre of an oil boom beginning in the 1870s (6), but the growing city struggled to gain sufficient water supplies for the burgeoning population and outbreaks of cholera were frequent. Having made several visits to the region, W.H. Lindley's proposal was to construct a 110 mile (177 km) pipeline to bring water from the Caucasus Mountains. He personally presented his plans to the Baku Duma (or parliament) on 1901 June 23 and was awarded the contract. The audacious project was technically immensely challenging. The pace was slowed by the political instability in Baku which resulted from the revolutionary events in Russia in 1905 to 1907. Nevertheless, the project was duly completed and to this day the pipeline carries water to central Baku. As a mark of gratitude and respect for completing such a monumental project, W.H. Lindley was granted honorary citizenship by the Baku Duma during his last visit to Baku in 1916 January; he died of a stroke in London in 1917 December.

W.H. Lindley's father, William Lindley (1808-1900; Figure 3), was also a famous civil engineer and they worked on several projects together (7). For example William the father designed the Warsaw waterworks between 1876 and 1878 and his son directed construction between 1881 and 1889. As a young engineer, William Lindley worked with other famous engineers of the age, including Marc Isambard Brunel (1769 – 1849), the father of Isambard Kingdom Brunel, and Francis Giles (8) (1787-1847). Early on in his career, William was involved with the construction of the Newcastle and Carlisle railway and the London to Southampton railway (9). In 1834 William went as Giles' assistant to survey the route for the Lűbeck to Hamburg railway and in 1838 he was commissioned to build the Hamburg-Bergedorf railway (10). Hamburg itself was destroyed by fire in May 1842 and William became a member of the committee responsible for rebuilding the city centre. In this role, he designed a new sewerage system for the city. In 1860, William moved to Hamburg with his three sons, William Heerlein Lindley, Robert Searles Lindley (born 1854) and Joseph Lindley (born 1859). He was involved in other civil engineering projects, including designing the sewerage system in Frankfurt-am-Main, where Max Lindley was later to be born. There is a statue of William Lindley in Hamburg (Figure 3) and streets are named after him (*Lindleystraβe*) in Hamburg, Warsaw and Frankfurt-am-Main. During the late 19th Century public health concerns, notably typhoid and cholera associated with rapidly growing urban populations, led many cities across Europe to commission water and sewerage systems. Consequently Lindley's designs were in demand across Europe, and many were family projects conducted together with his sons, including in Dűsseldorf, St. Petersburg, Budapest and Moscow (11). In this way the Lindley family name became well known throughout continental Europe.

Thus with his father and grandfather being prominent engineers, it was not surprising that Max Lindley should himself become an engineer, as will be discussed later. Lindley's great-grandfather, Joseph Lindley was, by contrast, an Assistant at the Royal Greenwich Observatory, so perhaps there was a vein of astronomy running in the family too (2).

**School and University**

Max Lindley lived at Frankfurt until the age of fifteen, when he was sent as a boarder to finish his education at Sherborne School in Dorset from Michaelmas term 1905 (12). There he exhibited strong musical talents, playing the violin and singing in several school concerts. He was described in the Sherborne School magazine as "the best violinist the School has had lately" (13) and in later years became an accomplished violinist (2). Music developed into a life-long interest and he played the violin in various Quartets throughout his life; he also developed a strong interest in early polyphonic choral work (14). While Lindley was at Sherborne he boarded at Harper House which was then run by Thomas Alfred Bell, M.A., an Exhibitioner of Trinity College Cambridge. Members of Harper House in 1907 can be seen in Figure 4. It was perhaps in connexion with Bell's contacts that Lindley went up to Trinity College Cambridge in 1910 where he read the Engineering Tripos, graduating BA in 1913, later becoming MA and Associate Member of the Institution of Civil Engineers (AMICE).

**First World War service**

Max Lindley volunteered for military service immediately the First World War broke out in August 1914, joining the Royal Engineers. He underwent basic training at Biggleswade and took a section of Engineers to the Western Front in November 1914, being gazetted as Second Lieutenant on 9 December 1914 (15). Just before he left England, he married Miss Florence Spencer whom he had known since school teenage days and they remained married until his death – they had no children (16). As a Brigade Signals Officer, Lindley was involved in setting up communications in the trenches; the part of the Royal Engineers of which he was a member later became the Royal Corps of Signals (17). Lindley served on the Western Front throughout the war, including at Ypres and the Somme. Although he rarely spoke about his experiences in the trenches, he did describe to his great niece in later years the many occasions he had to leave the trenches and crawl though barbed wire to lay and repair communications wires which allowed people in different parts of the front line to communicate with each other and with headquarters (14). Although sounding like acts of derring-do to the small girl, this was clearly a very hazardous operation. He was awarded the Military Cross in 1917 and was also mentioned in one of Sir Douglas Haig's despatches. A picture of Lindley taken during the war is shown in Figure 5.

Eventually the Armistice was signed and Lindley was demobilised in 1919 with the rank of Captain. He found the whole experience of fighting in the First World War deeply distressing as he knew that it was likely that he was fighting members of his family and his friends: he had lived in Germany until the age of 15 and of course his father and grandfather had lived there for many years, with several family members still residing in the country (14).

During his time in the trenches he developed a friendship with two members of Robert Falcon Scott's (1868-1912) last expedition to the Antarctic (18): Sir Charles Seymour Wright (1887-1975) and Sir Raymond Priestly (1886-1974). The three had much in common: all served in the Signals and were awarded the Military Cross. Lindley said: "C.S. Wright I knew quite well. He often came to me when I was a Brigade Signals Officer – I still have a memento of him in the form of a scrap of paper which he left behind one day – Christmas Day – when he came to my dugout on the Ypres Canal. That was 1915, Ypres" (19), which, in common with most British soldiers, he referred to as "Wipers" (14). Wright (20) (Figure 6) was a member of the team that set off with Scott from the base camp at Cape Evans on 1 November 1911 with the intention of reaching the South Pole. However, on 22 December at latitude 85º15' south, some 500 km from the pole, he was in the first supporting party which Scott sent back to base. Later, after Scott failed to return, he joined the 8-man search party, which on 12 November 1912 first spotted the tent containing the bodies of Scott, Edward Wilson and Henry Bowers, who had all perished on their return trek from the South Pole eight months earlier.

Priestly (21) (Figure 7) was in an Antarctic surveying expedition sent forth from the Cape Evans base camp and which on 3 February 1911 encountered Raold Amundsen's (1872-1928) ship, *Fram*, news of which they reported back to Scott. In late January 1912, during the Antarctic summer, Priestley embarked on another surveying mission, but his party encountered unseasonably bad weather and had to remain in an ice cave for 7 months during the following winter, eking out their 8 week rations with seal and penguin meat. They then endured a five week trek back to Cape Evans, eventually arriving on 7 November 1912, only then learning about Scott's death. On returning to England, Wright married Priestley's sister. Priestley was Lindley's Colonel at the time of his demobilisation.

Lindley also met a further member of Scott's expedition, Frank Debenham (22) (1883-1965; Figure 8) "in Stratton's Rooms at Cambridge" (2) – F.J.M. Stratton (1881-1960; Figure 9) was an astrophysicist who later became Professor of Astrophysics at Cambridge (23).

**Variable star observations and BAA Variable Star Section Secretary**

After Lindley was demobilised in 1919, he joined the British Thomson Houston Co. (24) in their electrical department at Rugby (25). However, he relinquished the position in 1923 for health reasons (possibly tuberculosis (14)) which led him to live in Switzerland for some time. Returning to England in 1924, he moved to Trevone, near Padstow, Cornwall. His house, Pentonwarra (Figure 10), was his home for the rest of his life and he delighted living there, devoting his time to his interests, just a few metres from the sea. Over the years he extended the house considerably.

Whilst at Rugby, Lindley joined the BAA, being elected on 28 January 1920. His membership was proposed by two eminent members of the BAA: Rev. T.E.R. Phillips and C.P. Butler. At the time, Rev. Theodore Evelyn Reece Phillips (1868-1942) was Director of the Jupiter Section of the BAA (which he Directed 1900–1933) and a recent past President of the BAA (1914 -1916); later he became Director of the Saturn Section from 1935–1940 (26). Although primarily a planetary observer, Phillips did make variable star observations. His 1916 Presidential address was on the analysis of the light curves of long period variables (LPVs). Lindley's other proposer, Charles Pritchard Butler (1871-1952) (27), was a solar astronomer, who worked for a time with Sir Norman Lockyer (1836-1920). Initially Butler was at the Solar Physics laboratory at South Kensington and then moved with it to Cambridge. In 1913 he was acting Director of the Kodaikanal Observatory in India.

Lindley became a Fellow of the RAS on 12 May 1922, his membership proposal being witnessed by F.J.M. Stratton, T.E.R. Phillips and J. Jackson (28). Dr. John Jackson (1887-1958) (29), CBE, FRS, a colleague of Stratton, was working at the time at the Royal Greenwich Observatory. He went on to be H.M. Astronomer at the Cape (South Africa) between 1933 and his retirement in 1950. He received the Gold Medal of the RAS in 1952 for his work on stellar parallax and proper motions and served as President from 1953 to 1955.

It has not been recorded what stimulated Lindley's interest in variable stars, but his first variable star observation, that of the long period variable (LPV) T Cas, was made on 10 January 1921 and he started contributing observations to the VSS the same year. This was the final year of service of then VSS Director, Charles Lewis Brook (1855-1939) (30), who was Director between 1910 and the end of 1921. In 1922 Félix de Roy (1883-1942; Figure 11) took over as the fourth Director and, since he was resident in Antwerp, Belgium, he appointed A.N. Brown as VSS Secretary essentially to run the Section on a day to day basis (31). De Roy organised the first ever VSS meeting on 25 October 1922 (32) in the Smoking Library at Sion College, London, which was then the venue of main BAA meetings. The meeting was timed so as to take place just before the Association's AGM and lasted 1 hour. Lindley attended the meeting, along with de Roy and six other members including W.H. Steavenson (1894-1975), Harold Thomson (1874-1962) and E.H. Collinson (1903-1990; Figure 12) (33).

The BAA Variable Star Section (VSS) database contains some 18,355 observations of 62 stars made by Lindley between 1921 and 1953 (34). Most of the stars were LPVs (35), which represented the bulk of the VSS target list at the time, although he observed other variables including novae, such as Nova Her 1934, DQ Her (36), and dwarf novae.  His most frequently observed star was the dwarf nova SS Cyg, with 1595 observations. Figure 13 presents Lindley's annual totals of observations and shows that his most productive years were the late 1920s and early 1930s; the year with the highest number of observations was 1927 with a total of 2499.

Most of Lindley's observations were made with his 5½ inch (140 mm) Cooke refractor, supplemented with 2 3/8 inch (60 mm) and 0.8 inch (20 mm) refractors (37). Lindley's first paper, published in the JBAA in May 1927, was on his variable star observing instrumentation (38). In the paper he noted that the 5½ inch telescope was equipped with a revolving eyepiece holder, made by Cooke, Troughton & Simms, which allowed 3 eyepieces to be exchanged quickly. He generally preferred Monocentric eyepieces when observing variables as they gave sharper stellar images and a blacker background than the standard Huyghenian eyepieces, which sometimes resulted in a gain of half a magnitude in limiting magnitude. The 2 3/8 inch telescope was also constructed by Cooke, Troughton & Simms according to his own specification. The objective was a triplet with a focal length of 9½ inches (240 mm), giving a focal ratio of f/4. It was specifically intended for variable star observing and Lindley noted that "[b]oth the light grasp and definition are a very remarkable advance on those of prismatic binoculars". He generally hand-held the instrument, often from a deckchair, and commented that the wide field of view "yields most beautiful views of the broader aspects of the Galaxy" (39). The main telescope was housed in an observatory in the garden at Pentonwarra (40)

The BAA VSS Secretary, A.N. Brown, died suddenly in November 1934. By that time Lindley was a well known member of the Section and a highly experienced observer, so it was natural for de Roy to appoint him as Brown's successor (41). He also had the necessary time available as he had not returned to work following his illness in 1924. As Secretary, Lindley was "to receive at his address all observations of Members of the Section residing in the British Empire", to distribute star charts, to maintain a record of members' original observations and to answer enquiries from new observers. Thus over the next few years, Lindley effectively took over the day to day running of the Section (42). This allowed de Roy to continue his important work of analysing the results of the Section. de Roy's *magnum opus* was the BAA Memoir on the Section's observations of LPVs between 1925 and 1929 (43), published in 1934. This volume contains more than 400 pages and contains 59938 observations of 51 LPVs, but de Roy credited Lindley as having "taken a large part in the preparation of [the] Memoir" (41). Lindley also contributed towards the cost of publishing the Memoir (44).

At de Roy's request, Lindley also undertook the annual analysis and reporting of four "irregular" variables covering the years 1927 to 1934, which he published as four papers in the JBAA (45) (46) (47) (48). The stars were U Gem, SS Aur (both now known to be dwarf novae), R CrB and R Sct (a pulsating star of the RV Tau family). In a further paper, Lindley presented a detailed analysis of the times of maxima of 19 LPVs which he undertook by reviewing light curves of the stars between 1900 and 1919 based on BAA VSS observations (49).

Variable star observations crucially depend on the availability of reliable comparison star sequences. It was not uncommon for sequences to deviate significantly, especially at the fainter end of telescopic variable sequences. The bulk of BAA VSS sequences were based on photometry from the Harvard College Observed published in 1900, 1902 and 1908 and by the mid 1930's de Roy and Lindley realised that revision, and extension to fainter magnitudes, was necessary. Furthermore some sequences, especially for novae, were derived by visual means by inspection of the field by experienced observers (50). The decision was therefore taken to update the VSS sequences with new photometry published at the end of 1935 by Prof. Samuel Alfred Mitchell (1874-1960) of the Leander McCormick Observatory at the University of Virginia, USA (51). Consequently a VSS Chart Committee was established to update the charts and sequences comprising Lindley, Steavenson, Patston, Holborn and Lane Hall (52). The first of the new draft charts were available by August 1938 and Lindley delighted in showing them to Mitchell at the IAU meeting in Stockholm that month (53). One thing that slowed the progress of the Chart Committee was sifting through the comparison stars that were suspected of variability as clearly these needed to be eliminated from the sequences. Of the 62 variable star charts that were to be prepared, Lindley thought that at least 55 of the original comparison stars were variable (53). Lindley was keen to check that Mitchell's comparison star magnitudes were consistent with what one actually sees in the eyepiece, rather than simply accepting the revised magnitudes at face value (54).

In early September 1932 Lindley and his wife travelled to the USA to attend the 4$^{th}$ General Assembly of the IAU in Cambridge, Massachusetts, in which de Roy also participated. Whilst there he visited the headquarters of the American Association of Variable Star Observers (AAVSO), which was located at Harvard College Observatory in Cambridge (55). There he met the AAVSO's founder, William Tyler Olcott (1873-1936; Figure 14), Leon Campbell (1881-1951; Figure 15), who was Recorder of the AAVSO and with whom he corresponded during subsequent years, and Harlow Shapley (1885-1972; Figure 16), Director of Harvard College Observatory. A photograph of some of the delegates is shown in Figure 17. In later years, Lindley attended further IAU General Assemblies: the 5$^{th}$ in Paris in 1935, the 6$^{th}$ in Stockholm in 1938 as mentioned above and the 9$^{th}$ in Dublin in 1955 (56). Naturally enough, he was especially interested in the discussions at IAU Commission 27 on Variable Stars.

Even before de Roy's and Lindley's visit to AAVSO Headquarters, there was already a strong link between the BAA VSS and the AAVSO. Both Charles Lewis Brook and Félix de Roy had regular correspondence with Campbell. Lindley decided to join the AAVSO more or less as soon as he returned to England (57) and Campbell asked him if he might observe several stars that were on the AAVSO programme, but not the BAA's (58). Lindley purchased a large batch of AAVSO charts with the aim of comparing with, and updating, the BAA's charts by the Chart Committee (59). Lindley and Campbell also corresponded on the subject of stars in the fields of known variables, and which had been used as comparison stars, but which BAA VSS members suspected of being variable (60).

In 1936 Lindley travelled to the Greek Island of Chios to observe the total eclipse of the sun on 19 June, along with other members of the BAA. He presented his observations at the October 1936 meeting of the Association, noting "the eclipse was exceedingly well seen and the people of the island had been most hospitable and helpful" (61).On the same expedition was Dr. R.L. (Reggie) Waterfield (1900-1986; Figure 18) who became a good friend of Lindley. Waterfield was Director of the BAA Mars Section from 1932 to 1942 and an accomplished astrophotographer, especially of comets. He was a frequent visitor to Pentonwarra and when later he became confined to a wheelchair as a result of poliomyelitis, he found his stays there quite convenient as much of the property was on a single level. Waterfield shared

Lindley's love of music, he being an accomplished pianist, and they played together (14). Waterfield was BAA President from 1954 to 1956.

**Director of the Variable Star Section**

Félix de Roy was able to devote increasingly less time to his responsibilities in the late 1930's due to his responsibilities as Editor of an Antwerp newspaper, as well as due to bouts of poor health. There were also concerns about the looming threat of a new war. He had been ill during the first few months of 1939 and with the outbreak of war in September, contact with him was lost. Lindley placed a special announcement in the October 1939 Journal in his capacity of VSS Secretary stating:

> "*As the Director [de Roy], owing to present difficulties in communication, is unable himself to make such a request, the Secretary asks all members of the Section to do their utmost to keep the stars on their programme under observation, so that our long series may not be interrupted. During these years every single observation will be of special value*" (62)

Given that Lindley had been carrying a large part of the burden of running the Section for several years, he was naturally invited to become the next Director by BAA Council and this was announced in the November 1939 Journal. Lindley stated (63) that he would address three immediate tasks:

(1) to bring and keep up to date the Interim Reports on the work of the Section
(2) to revise the comparison star sequences and the charts; and
(3) to compile a Memoir on Long Period Variables in 1930-1934

The number of VSS publications had decreased during the later years of de Roy's Directorship and by the mid 1930s very little was appearing in the Journal. de Roy had published annual reports in the JBAA on LPVs covering the years 1921 to 1930, as well as several on Irregular Variables and his *magnum opus* was the BAA Memoir on the Section's observations of LPVs between 1925 and 1929 (64). de Roy had started work on the next Memoir, covering LPVs in 1930 – 1934, as far back as 1936 (52), but little progress had been made by the time Lindley took over as Director. As we have seen, de Roy had requested Lindley to continue the Journal series on Irregular Variables, which continued until 1934. Furthermore de Roy also requested his predecessor as Director, Charles Lewis Brook to prepare annual publications on the dwarf nova SS Cyg, which continued until 1927 (30). It therefore fell to Lindley to reinvigorate the VSS reporting.

However, there can hardly have been a less opportune time for Lindley to take over as Director. With the outbreak of war, Lindley knew it was only a matter of time until he would be called up on active service as he was still on the Reserve of Officers. He told the BAA President, B.M. Peek (1891-1965), that he would endeavour to maintain the Section correspondence for as long as possible (65), but it was inevitable that VSS responsibilities would take a back seat. Lindley lamented in mid

September 1939 that although he had been looking forward to a busy winter of VSS activities, the onset of war would necessarily restrict these (65). He pointed out in one of his letters to Leon Campbell that one advantage of the remoteness of his Cornwall residence was that VSS records and other documentation would be relatively safe there from enemy action (65).

Lindley's call up was not long in coming, his orders arriving at Trevone on 8 February 1940, giving him one week to make his final preparations for departure, which including informing VSS members (66). A photograph of Lindley in uniform taken during the war in the garden at Pentonwarra, overlooking the sea, is shown in Figure 19. He was posted to the Royal Corps of Signals camp at Catterick in Yorkshire, where he was responsible for training recruits. Evidently he found the work interesting and he enjoyed walking in the surrounding countryside, including the moors and dales. He maintained correspondence with VSS members from his Catterick base and undertook some variable star observations whilst there. In such a letter to his good friend and fellow variable star enthusiast, E.H. Collinson in March 1940, he commented on the importance of his hobby to him (67):

> *"What a sensible entertainment this astronomy of ours is to be sure. Present circumstances and occupations bring that out very clearly. I suppose however we might quite rightly say that we are fighting for an opportunity of continuing with peaceful observation. What a mess we have got ourselves into!"*

Later he spent a long time in hospital in Catterick and was invalided out of the army with the rank of Major. He returned home to Trevone, according to his obituary by Patston, "rather disenchanted". Lindley himself noted that "these are not days for people a large proportion of whose lives were spent pre-1914" (2).

As Colin Munford noted in his review of VSS activities written for the BAA Memoir on the Association's second fifty years (68), the numbers of VSS observations were greatly reduced during the war as so many people were involved in war duties, but variable star estimates were received from unusual places such as El Alamein and Mersa Matruh during the Western Desert campaign. Lindley reported that a total of 11,148 observations were submitted to the VSS in 1939, there were 2559 submitted in 1940 and 1703 for 1941 (69). Following the end of hostilities, there was an increase in observational activity with about 13,000 observations being submitted to the VSS in 1946. However, the renewed activity was not sustained. Moreover, few Section reports were written, which also gave the impression of a rather moribund Section. Nevertheless during the 1952-53 session 11 new members joined the VSS (70).

Lindley's energies in the 1950s were directed towards continuing to update the comparison star sequences on VSS charts and to producing the long awaited Memoir on VSS observations of LPVs in the years 1930-1934. He made few of his own variable star observations in the 1950s (Figure 13). The Preface to the Memoir was dated by Lindley as November 1957 and the final draft was reviewed by VSS members, including Lindley, at the BAA meeting of 27 November 1957 (71). The Memoir finally appeared in 1958 (72) and contained some 50,000 observations of 51 LPVs made by 38 observers. To reduce the cost of publication, the Memoir was issued on microfilm, with only the Preface being printed (73). However, this new method of reproduction was not uniformly welcomed by VSS members, some of whom preferred that it should still be printed in its entirety (74). The small number of copies, and the fact that a microfilm reader was required to view the content, inevitably meant that the Memoir did not receive the attention it might otherwise have done. Lindley noted in the Preface that he received considerable assistance in preparing the Memoir from other Section members, notably P. Harvey (75), as well as H.H. Hammond, F.M. Holborn (1884-1962; Figure 21) and G.E. Patston (Figures 20 and 22). Lindley appointed P. Harvey and J. Friends as "assistant to the Director" in 1954 (76).

Almost as soon as the Memoir was complete, discussions began amongst BAA Council members about the need to find a successor to Lindley as Director, with the aim of reinvigorating the Section. In spite of completing the Memoir, there was concern amongst VSS membership and some Council members that insufficient attention had been paid to publishing the work of the Section in the Journal in recent years and there was a growing feeling that a new Director should be recruited to take this matter forward. Lindley was asked on several occasions to make suggestions as to his successor although he was reluctant to make any formal proposal. Nevertheless, his personal preference was E.H. Collinson (77), whom he had known and respected for many years. The main problem was that Collinson was already Director of the Mars Section, a position which he did not wish to relinquish. Collinson was popular with VSS members and he was approached by several people, including G.E. Patston and F.M. Holborn, and asked if he would consider taking on the VSS Directorship, but he politely declined. Patston's letter to Collinson (78) in February 1958 sums up the sense of urgency to find a new Director:

> *"At the present time I am feeling very far from happy over the poor old VSS and the treatment meted out to its members. I wish with all my heart that you could be prevailed upon to hand your planet [Mars] over to somebody – anybody you like (!) – and take the helm of the VSS if the present helmsman relinquishes it, which I consider far from unlikely."*

In early April 1958 Lindley himself formally asked Collinson if he could propose Collinson to BAA Council for the position of Director (79):

> *"May I therefore ask you whether you would agree to my putting your name forward as willing to shoulder this heavy and important burden? My eyes are entirely open as to what I am asking of you and I only do it in the interests of the Section. You have all the necessary qualifications and it is only a question of whether you can find it in yourself to do it. I know no one whom I would*

> *rather see in my shoes and of that there is no doubt. And I should of course give you all possible assistance as long as you require it".*

In reply (80), Collinson explained that he had given the proposal much thought, but concluded that he simply didn't have the time to direct two BAA Sections, the VSS and the Mars Section, and much of his time was taken up by professional activities as an Ipswich solicitor and by family responsibilities. Collinson suggested that Harvey (who prepared much of the 1958 VSS Memoir with Lindley) and Andrews should be considered for the post.

E.A. Beet (1904-1997), who was BAA Secretary at the time, visited Pentonwarra to explain the BAA Council's desire that a new Director should be found promptly (81). At its next meeting, in April 1958, Council set up a small committee comprising Lindley, Collinson and Holborn to progress the appointment of a new Director (82).

Holborn himself, a very active and long-standing member of the VSS having joined in 1924 (83), had clear views about the attributes of a future Director:

> *"Also I have said that perfection is unattainable, and that a hot stuff scientist at the head of the VSS is less important than getting things going again quickly, for once observations are published, they are available for scientists to play about with if they want to"* (84)

Patston held a similar opinion, saying that he preferred that the new Director should not be:

> *"some splendidly flamboyant youngster...We older members of the VSS who look back with ever increasing nostalgia on days of great activity and greater happiness (greater illusion?) now look for someone mature, of our own age and standing, and experience of Directing"* (78)

Holborn suggested to Lindley that R.G. Andrews would make a suitable Director, having first ascertained that Andrews would be interested in the job (82), but initially the idea was rejected by Lindley (85). Meanwhile some VSS and Council members, still frustrated by lack of recent VSS publications in the Journal, suggested that a group be appointed by Council, with Lindley as convenor, to oversee the preparation of publications of the Section with the aim of bringing them up to date (86). Lindley viewed this as a challenge to his authority, thereby precipitating his resignation as Director. With Collinson ruling himself out as successor, and there being no other obvious candidate, the committee that had been set up to look into who should become the new Director moved forward with the proposal to appoint Andrews and this was accepted by BAA Council.

Reginald Gordon Andrews (1903-1996; Figure 22) had joined the association in 1945 and began contributing observations to the VSS soon after. In the Section Report for 1946, Lindley commented on the good contribution of Andrews of 600 observations made with a 2½ inch (65 mm) refractor from his residence in Sussex (87). Thus Andrews had the necessary practical experience in variable stars to take on the Directorship. Andrews stayed at Lindley's residence for 6 days to learn the responsibilities of being Director, after which he commented (88):

> "I feel I have a good grasp of what the job entails. It now remains to harness the motive power to the machinery!"

Andrews immediately set about preparing reports for publication and the first of 33 Section reports appeared in the Journal in 1959. Colin Munford noted that "[t]his hard work and enthusiasm attracted observers and the Section steadily grew in numbers and observations reported" (87). Andrews continued in office until 1964 when he resigned following a dispute with BAA Council about the publication of VSS reports in the Journal (89).

After Lindley had handed over the reins to Andrews, he "generously presented his entire library including the B[onner] D[urchmusterung] Catalogue and charts, atlases and Harvard volumes to the Association" (2).

Lindley launched at least one person on a future astronomical career. Prof. Jeremy Tatum (retired Professor of Physics and Astronomy at the University of Victoria, British Columbia Canada) was living in Trevone when:

> "one evening (probably aged 13 [probably in 1948]) I happened to notice a very bright star, and, out of idle curiosity I looked at it through my father's binoculars. To my astonishment I saw that it was accompanied by four satellites in a straight line...... My parents said that there was an astronomer living in a house overlooking the beach, and I should go down and tell him about it. That was Mr Lindley, and my life's course was determined from that day. Mr Lindley told me that it was in fact Jupiter (90)........ and he started telling me all sorts of exciting things about astronomy. He even showed me his huge (all five-and-a half inches of it) telescope at the bottom of his garden, and he asked me if I would like to look through it at night. You just cannot begin to imagine my excitement when I first looked through a telescope" (91).

A friendship soon developed as Tatum visited Lindley on a daily and nightly basis. Lindley told him about his astronomical heroes, William Herschel and F.W. Argelander; they listened to classical music records (92) and took frequent walks. They looked at many objects through Lindley's telescope, including some variable stars, although Tatum was not bitten by the variable star bug (93). Tatum also recalls at least one of Waterfield's visits to Pentonwarra (94). As Tatum's interest in astronomy grew:

> "Mr Lindley lent me many astronomy books to read. I would usually devour them in a few days and then go back and ask to borrow more. I suppose it eventually reached the stage when I was becoming a bit too enthusiastic, so he lent me Smart's Spherical Astronomy, presumably to shut me up. When I

> *opened it up, I saw that it was full of sines and cosines, which I had heard about in school but wasn't in the least interested in - until I opened that book and found that astronomers actually used sines and cosines. I was fascinated, and immediately sines and cosines became a focus of intense interest! It took me more than a few days to finish that book, but I can assure you that I read it from cover to cover, and I still refer to it from time to time"*

Lindley ended up giving the young Tatum several astronomy books, including Barnard's "Photographic Atlas of Selected Regions of the Milky Way" (95), the "Collected Works and Correspondence of Sir William Herschel" and a biography of Sir Arthur Eddington (96). Tatum's life course was thus set:

> *"It was during my acquaintance with Mr Lindley that I decided that I was going to be an astronomer. Nobody took me seriously. It was all very well to keep astronomy as a hobby, but what was I going to \*do\*? Even Mr Lindley warned me that I'd have to be very good at maths. Very well, then, if that's what it meant, I would be very good at maths. Whatever it took, but I was going to be an astronomer".*

Tatum went on to take a degree in Physics at the University of Bristol (1957) and a PhD from the University of London (1960). He recalls fondly the encouragement he received from Lindley and notes that:

> *"Mr Lindley was a rare and old-fashioned type. Some might have called him "posh", not in any derogatory sense, but only to recognize that he was an exceptionally well-spoken and cultured gentleman of the old school. When one of the old cottagers fell ill, Mr Lindley would walk regularly the full length of the village (i.e. not just when he happened to be passing) to inquire with great sincerity how old Mr Tresomebody was".*

Although Lindley might not have achieved all he had set out to do when he became Director in terms of preparing publications on VSS observations, partly due to disruption caused by war service, he was nevertheless welcoming towards new variable star observers who contacted him. One such was Colin Munford (97) (Figure 23) who joined the VSS in 1953. Munford recalls receiving 4 letters from Lindley, which were very encouraging to the new variable star observer and contained constructive criticism of his early results along with detailed advice, which was highly appreciated by Munford, allowing him to improve his observations (98) and to become a very productive observer over a period of many years. Gordon Patston had a similar experience when he was taking his first steps in observing variable stars in 1934. He contacted Lindley who was then in his first year as VSS Secretary:

> *"[Lindley], then always approachable, patient and sympathetic, somehow found time for reams of encouragement and guidance".* (2)

Patston soon became a very active and reliable member of the VSS, observing with a 12 inch (30 cm) Newtonian at his home in Streatham (99). Patston also had many other pleasant memories of an enthusiastic Lindley during the 1930's:

> "*It is good to recall those days, to summon up remembrance of things past: of last minute telegrams, "Lunch today"; of the "Catsmeat Club", the uneuphonious name for supper, where tables would late be littered with papers and charts, and where five whole minutes in private conclave with the harassed Secretary was a wonderful concession!; of reply postcards (penny postage then!) scribbled on the way home of something forgotten or of further advice needed. He was truly a wonderful correspondent then*". (2)

Patston contrasts these pre-Second World War memories of a lively and gregarious Lindley with his later years as Director, clearly affected by the war and being "rather disenchanted" (2). In the post-war years, Lindley became somewhat reclusive and preferred not to venture far from home. He was entirely happy at Pentonwarra, pursuing his lifelong interests of astronomy and music, although he enjoyed walking along the coastal cliffs or across the fields into Padstow to do the shopping (100). One school of thought is that the cause of Lindley's disenchantment, referred to in Patston's obituary, was his despondency over man's inhumanity to man as exemplified by the onset of a second World War and this might have caused him to become more introspective in later life. As noted above, he was clearly upset by having to fight, as he saw it, his friends and family in the First World War and now he saw the tragedy repeating itself (14) . However, Lindley should be remembered for his stalwart service to the BAA VSS over a period of several decades as an observer, as an author of reports and papers, as a leading member of the Chart Committee, as Secretary and as Director. And his *magnum opus* will always be his Memoir on LPVs. In later years, after his time as Director, Lindley was invited to become President of the BAA, but declined as "by now he found the distance too great for frequent attendance and felt he might not do justice to the honour" (2). He spent the rest of his life with his wife at his beloved Pentonwarra, passing away on 2 September 1972.

**References and notes**

1. Toone J., JBAA, 120, 135 (2010). John Toone's excellent and highly readable account of "British variable star associations, 1848-1908" describes the origin and evolution of the BAA VSS

2. Patston G.E., JBAA, 83, 201-203 (1973)

3. http://en.wikipedia.org/wiki/William_Heerlein_Lindley

4. At the time Baku was part of the Russian Empire

5. Zelichowsky R., Azerbaijan International, Summer 2002, "Water- Not a Drop to Drink, How Baku Got Its Water-The British Link - William H. Lindley". Available online at

http://azer.com/aiweb/categories/magazine/ai102_folder/102_articles/102_shollar_zelichowski.html

6. Large-scale oil exploration started in Baku in 1872, when Russian imperial authorities auctioned parcels of oil-rich land around Baku to private investors, including the Rothschilds. By the beginning of the 20th century almost half of the world's oil was being extracted in Baku

7. http://en.wikipedia.org/wiki/William_Lindley

8. Francis Giles was initially a canal engineer and surveyor and later became a railway engineer

9. Skempton, A. W., et al., eds. Biographical Dictionary of Civil Engineers, Volume 1, 1500-1830 (2002).London: Thomas Telford on behalf of the Institution of Civil Engineers

10. The Hamburg-Bergedorf railway, opened in 1842, is one of the oldest lines in Germany and was the first railway line in Northern Germany

11. William Lindley was asked to design a sewerage system for Sydney, Australia, but he turned it down as he had recently been commissioned to proceed with a project in Warsaw and thus did not have time available for the new project

12. Rachel Hassall, School Archivist, Sherborne School, Personal Communication (2010)

13. The Shirburnian, November 1907 (magazine of Sherborne School). The article goes on to say "Lindley's playing of a berceuse of Godard's was excellent. We seemed to see the nymph sleeking her soft alluring locks". References to Lindley's concert performances also appear in the Shirburnian in July 1905, November 1906 and June 1907

14. Morwyn Porter, personal communication (2010), Mrs. Porter is Lindley's great niece and lived in Pentonwarra with Lindley and his wife as a child in the 1940s and 50s

15. The London Gazette, 12 February 1915

16. Florence died in 1975

17. A Royal Warrant for the creation of a Corps of Signals was signed by the Secretary of State for War, Winston Churchill, on 28 June 1920. Six weeks later, King George V conferred the title of Royal Corps of Signals. The Royal Signals, as it is commonly called, still provides the battlefield communications and information systems essential to all operations of the British Army

18. Sometimes known as the "Terra Nova" Expedition, after Scott's ship

19. Lindley's quotation appears in his JBAA obituary written by Patston and is from a personal letter from Lindley to Patston

20. Wright, nicknamed "Silas", was born in Toronto, Canada and took a degree in Physics from the University of Toronto. Between 1908 and 1910 he did postgraduate research at the Cavendish Laboratory, Cambridge. During WW1 he helped develop trench wireless. Much of his career was spent in the Admiralty Research Department and in the Second World War he was involved in radar development. He was knighted for this work in 1946

21. R.E. Priestley was born in Tewkesbury and later studied geology at what is now Bristol University. After WW2 he conducted research on glaciers at Cambridge and in 1920 founded, together with Antarctic explorer, Frank Debenham, the Scott Polar Research Institute. He held a number of academic and administrative posts in England and Australia, including being Vice-Chancellor of Melbourne and Birmingham Universities. He was knighted in 1949

22. Frank Debenham, OBE, undertook surveying and mapping activities in Antarctica and worked closely with C.S. Wright on this task. After WW1 he co-founded the Scott Polar Research Institute with Raymond Priestley. In 1931 he was appointed Professor of Geography at Cambridge University. He was Vice-President of the Royal Geographical Society (1951-1953). He did not join Scott's final expedition due to a knee injury sustained in a football match in the snow

23. Frederick John Marrian Stratton, FRS, worked at the Solar Physics Observatory, Cambridge, where he was Assistant Director from 1913 to 1919, then Director from 1928 to 1946, then on the amalgamation with the Cambridge Observatory, he became Director of the Combined Observatories from 1946 to 1947. He was Professor of Astrophysics at the University of Cambridge from 1928 to 1947. In 1947 he was made a Fellow of the Royal Society

24. The British Thomson Houston Company was an electrical engineering company founded in 1894 as the British arm of an American parent, which then became merged into General Electric Inc. (GE)

25. Whilst at British Thomson Houston, Lindley and his wife lived at 26 Park Road, Rugby

26. Rev. T.E.R Phillips' RAS Obituary: MNRAS, 103, 70 (1943)

27. C.P. Butler's RAS Obituary: MNRAS, 113, 294 (1953)

28. RAS Fellowship proposal form signed 10 May 1922, elected 12 May. RAS Library - copy kindly provided by Peter Hingley, RAS Librarian

29. Obituaries of John Jackson: Spencer Jones H., MNRAS, 119, 345-348 (1959) and Monthly Notices of the Astronomical Society of Southern Africa, 17, 129 (1958)

30. For a description of the life and work of C.L. Brook, see Shears J., JBAA, accepted for publication (2010)

31. For a description of the life and work of F. de Roy, see Shears J., JBAA, accepted for publication (2010)

32. de Roy F., BAA VSS Circular 4 (1923)

33. Lindley was absent from the following VSS meeting of 31 October1923 (de Roy F., BAA VSS Circular 6, 1924), but he was present at the subsequent one on 26 November 1924 (de Roy F., BAA VSS Circular 7). Both were held at Sion College, London

34. Data from the BAA VSS database. This is likely to be an underestimate as it is believed that not all observations have yet been entered

35. LPVs are pulsating red giants or supergiants with periods ranging from 30-1000 days. There are two subclasses: Miras (named after Mira Ceti) and Semiregulars

36. The nova was discovered by Manning Prentice, then Director of the BAA Meteor Section, on the night of December 12 1934. See Prentice J.P.M., JBAA, 45, 120 (1935). It peaked 9 days later at magnitude 1.5. A slow fade followed, with the nova losing 3 magnitudes in 94 days, followed by a more rapid decline of 8 magnitudes in just one month. DQ Her then rebrightened to a second maximum of magnitude 6.5, before a slow fade to minimum. Further information about DQ Her, written by Gary Poyner, can be found on the VSS web site at http://www.britastro.org/vss/00191a.html, where a light curve can be viewed

37. Lindley noted that the 2 3/8" with a power of X8 gave a field of view of 6 1/2 degrees and the 0.8" a field of view of 18.5 degrees at X1.75

38. Lindley W.M., JBAA, 37, 264-265 (1927)

39. The current whereabouts of Lindley's telescopes in not known to the author, who would be interested to learn from anyone who has any information about them. The pier for the telescope still exists in the garden at Pentonwarra

40. Morwyn Porter recalls that the observatory had a rotating top with shutters that opened and it appears to have been of the traditional Romsey design

41. de Roy F., BAA VSS Circular 10 (1934)

42. Kelly H.L., "Variable Star Section", in "The BAA - The First Fifty Years", BAA Memoir 42, part 1 (1989)

43. de Roy F., Mem. Brit. Astron. Assoc., 31 (1934)

44. Report of the BAA meeting of 24 April 1929, JBAA, 39, 230 (1929)

45. Lindley W.M., JBAA, 44, 330-336 (1934)

46. Lindley W.M., JBAA, 45, 28-34 (1934)

47. Lindley W.M., JBAA, 46, 182-187 (1936)

48. Lindley W.M., JBAA, 49, 22-28 (1938)

49. Lindley W.M., JBAA, 43, 125-130 (1933)

50. An example is the sequence for HR Lyr (Nova Lyr 1919). Until recently, the AAVSO chart for HR Lyr was in part based on comparison stars whose brightness was determined by visual estimation by W.H. Steavenson

51. Lindley W.M., JBAA, 46, 223-225 (1936)

52. de Roy F., JBAA 46, 364 (1936)

53. Lindley W.M., Letter to Leon Campbell 20 May 1939, AAVSO Archive

54. Lindley discussed the magnitudes assigned by Mitchell with E.H. Collinson, another experienced BAA VSS observer, in 3 letters he wrote to Collinson on 11 February, 7 March and 15 March 1936

55. A total solar eclipse took place in August 1932, which de Roy observed from Maine, USA. I have no record that Lindley observed it, but given the fact that it took place so close in time and distance to the IAU meeting it seems likely that he did. Many observers, including the BAA's R.L. Waterfield, were unable to view the eclipse because of cloud or mist de Roy rode in a helicopter above the mist and had a good view. This is the first time that a solar eclipse was observed from a helicopter

56. Lindley attended the 1955 IAU in Dublin in his role as Director of the BAA VSS. Discussions at this IAU led him to conclude that a wider variety of variable stars should be included in the VSS programme, beyond the core of LPVs, such as Flare Stars

57. Lindley W.M., Letter to William Tyler Olcott (AAVSO), 8 November 1932, AAVSO Archive

58. Campbell L., Letter to W.M. Lindley, 25 September 1935, AAVSO Archive

59. Lindley W.M., Letter to Leon Campbell, 20 May 1939, AAVSO Archive. In reply Campbell agreed to Lindley's request for 244 copies of charts: Campbell L.C., Letter to Lindley, 5 June 1939, AAVSO Archive

60. Lindley W.M., Letter to Campbell, 23 May 1939. Campbell L.C., Letter to Lindley, 8 June 1939. Both letters from the AAVSO Archive. In addition, Lindley wrote to E.H. Collinson to ask his view on the suspected variability of comparisons in the field of. S Cas, R Cyg and V Cyg; Lindley W.M., Letter to E.H. Collinson, 3 May 1938

61. Lindley W.M., JBAA, 47, 15 (1936). In the report of the October 1936 meeting of the BAA

62. Lindley W.M., JBAA 49, 421 (1939)

63. Lindley W.M., JBAA, 50, 55 (1939)

64. de Roy F., Mem. Brit. Astron. Assoc., 31 (1934)

65. Lindley W.M., Letter to L. Campbell, 18 September 1939, AAVSO Archive

66. Lindley W.M., Letter to E.H. Collinson, 8 February 1940

67. Lindley W.M., Letter to E.H. Collinson, 1940 March 26. Letter sent from Catterick Camp

68. Munford C.R., "Variable Star Section" in Mem. Brit. Astron. Assoc., 42(2): The British Astronomical Association, the Second Fifty Years, R. McKim, ed.

69. Lindley W.M., JBAA 52, 309 (1942). The numbers appear in Lindley's annual report for the VSS for the 1941-42 session submitted to the BAA Council. He issued no annual report in 1940 and 1941 due to being tied up with military service. However, the current BAA database lists 8638, 5742, 1294 for the same years

70. Lindley W.M., JBAA, 63, 300 (1953)

71. Patston G.E., Letter to E.H. Collinson, 18 November 1957

72. Lindley W.M., Mem. Brit. Astron. Assoc., 38, 1958

publish all the VSS reports prepared by Andrews. This was ironic as the very thing that Andrews was asked to do when he was appointed Director was to bring the Section publications in the Journal up to date. It was because of Hyde's unwillingness to publish this material that Andrews resigned

90. Tatum consulted Pears' Cyclopaedia and discovered that Uranus had four satellites, so he assumed the object must be Uranus. Lindley was soon to put him right!

91. Tatum J., Personal communication (2011)

92. Lindley had a collection of 33 rpm records, which was rather a new format (78 rpm records existed before that). Lindley played Mozart's Clarinet Quintet, Elgar's Dream of Gerontius, Strauss' opera Die Fledermaus and pieces by Poulenc (which the young Tatum did not particularly enjoy)

93. Tatum comments: "I was never an active variable star observer, or indeed a serious amateur observer, though Mr Lindley certainly showed me many variable stars, and I still remember and look at the naked eye ones when I get the opportunity........ Mr. Lindley promised me a present if ever I was not able to see R Coronae Borealis. He showed me the method that he used to make magnitude estimates, by mentally bracketing the variable into tenths between two comparison stars"

94. Tatum recalls he "did meet Dr Waterfield while he was visiting Mr Lindley. He was confined to a wheelchair. They were both tickled pink that I, a mere schoolboy, had spotted some trivial mistake in one of Dr. Waterfield's books"

95. Tatum took the Atlas with him to the University of Victoria where he kept it in the chart room. In a reorganisation the University Library officially took possession of all the charts including the Atlas. It was only later that Tatum realised the value of the rare book, but by then it was too late: it had already become official University property. Tatum received a tax receipt for his "generous donation" and Lindley's book still resides in the University library; Tatum J., Personal communication (2011)

96. Sadly, Tatum's father disposed of some of the books that Lindley had given the young Jeremy. This happened whilst Tatum was away at Bristol University, his father assuming that because they were old, and because he would have access to the latest books at the University, he would no longer need them. He still has "biography of Eddington by the Canadian astronomer Vibert Douglas. I happened to refer to it recently and I see that it still bears a few small pencilled annotations in it made by Mr Lindley"

97. Colin Munford was awarded the BAA Steavenson Award in 1980 for his contributions to the Variable Star Section," both as an observer and as an active administrator". See JBAA, 90, 408 (1980)

98. Munford C.R., personal communication (2010). The first letter Munford received (24 Aug 1953) contained advice on locating a variable, W And - often one of the biggest challenges for a new variable star observer is finding the variable in the first place. Munford also recalls meeting Lindley briefly on two occasions at Burlington House in 1955 or 56. This took place following a Council meeting that Lindley had attended and before the main BAA meeting

99. Patston's obituary can be read in the Journal: Munford C.R., JBAA, 100, 246 (1990). Patston's Streatham Observatory is described in JBAA, 50, 216 (1940). After he retired he moved the observatory to Eastbourne. Patston served as VSS Secretary 1966-1973

100. Jeremy Tatum notes that Lindley "didn't have a car, so that he'd have to walk to Padstow for the shopping. There was one small shop in the village, and you could get veggies from the farms, but I do know that Mr Lindley often walked to Padstow. ...[sometimes] along the cliffs via Tregudda Gorge, which was a real, real long walk"

101. This photograph is used in: Kelly H.L., "Variable Star Section", in "The BAA - The First Fifty Years", BAA Memoir 42, part 1 (1989)

102. Pickard R.D., Personal communication; data from the BAA VSS database (2010)

103. The complete photograph, including identifications, can be seen in Popular Astronomy, 40, 453-459 (1932). A higher resolution image is included in Blaauw A., "The History of the IAU", publ. for the International Astronomical Union by Kluwer (1994)

**Acknowledgements**

I am most grateful for the assistance I have received from many people during the research for this paper. Morwyn Porter, Lindley's great niece, Colin Munford, a long-standing member of the BAA VSS, and Professor Jeremy Tatum, who as a boy lived near Lindley, kindly provided their firsthand knowledge of Lindley's life, character and activities. Richard McKim was helpful in many matters, especially concerning details of the history of the Association, as well as providing me copies of correspondence between E.H. Collinson and R.G. Andrews, F.M. Holborn & Gordon Patston and the photo of Lindley published in this paper. Both Colin Munford and Richard McKim kindly commented on a draft of this paper. Mike Saladyga (AAVSO) provided copies of letters between Lindley and Leon Campbell & William Tyler Olcott from the AAVSO archives, as well as providing the photograph of Olcott reproduced here. Roger Pickard provided details of Lindley's BAA VSS observations from the VSS database as well as a copy of the Preface to Lindley's Memoirs on LPVs; he also answered numerous enquiries from me regarding various historical details of the VSS. Rachel Hassall (School Archivist, Sherborne School) searched the Sherborne School archives and provided information on Lindley's school days, including copies of the School magazine and photographs. Claire Ray of Unique Home Stays (www.uniquehomestays.com), the company which lets Pentonwarra, Lindley's home for most of his life, gave me permission to use the photograph of Pentonwarra and also put me in touch with Morwyn Porter. Sheridan Williams has reduced the number of trips that I needed to make to the RAS Library in London to consult back numbers of BAA Journals, by diligently scanning the Journals and making them available on line to Members (a truly wonderful resource) – and he fast-tracked the availability of some specific editions to assist my research. Richard Baum provided encouragement to pursue this research and guidance on how to weigh the importance of different historical perspectives by considering how matters were viewed at the time that they actually happened, not just from the present perspective. He also provided insight into a number of matters concerning the history of the VSS, as did Storm Dunlop, John Toone and John Isles. Peter Hingley (RAS) looked after me on numerous visits to the RAS library and provided details of Lindley's RAS Fellowship application. This research made use of the NASA/Smithsonian Astrophysics Data System. Finally I thank the referees whose constructive comments have improved the paper.

**Address**

"Pemberton", School Lane, Bunbury, Tarporley, Cheshire, CW6 9NR, UK [bunburyobservatory@hotmail.com]

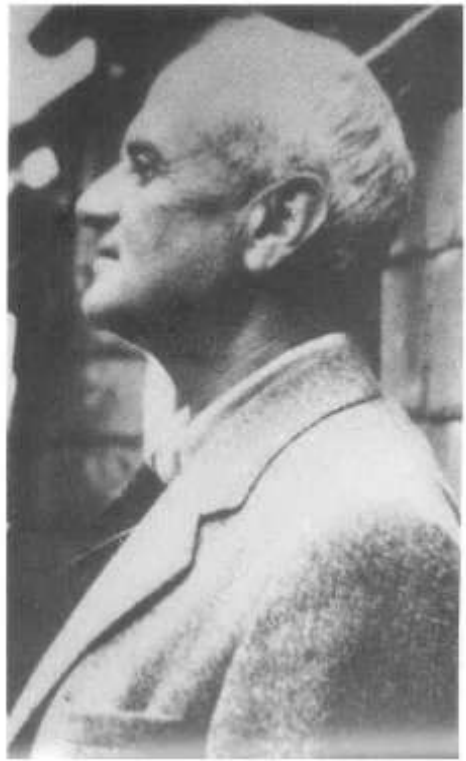

Figure 1: William Maximilian Lindley MC, MA, FRAS, AMICE (1891–1972) (101)

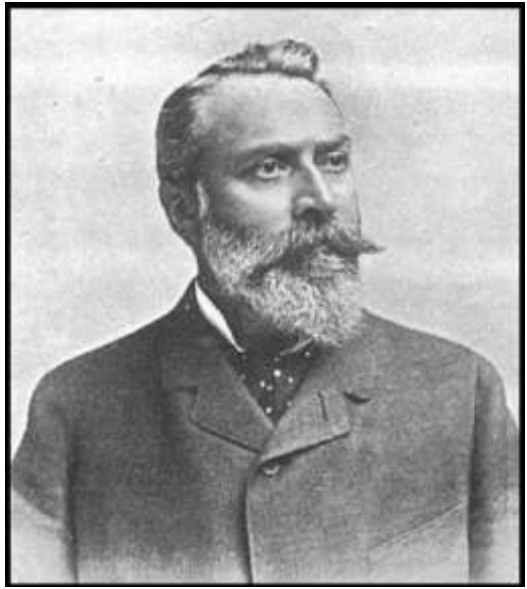

Figure 2: Sir William Heerlein Lindley (1853-1917)

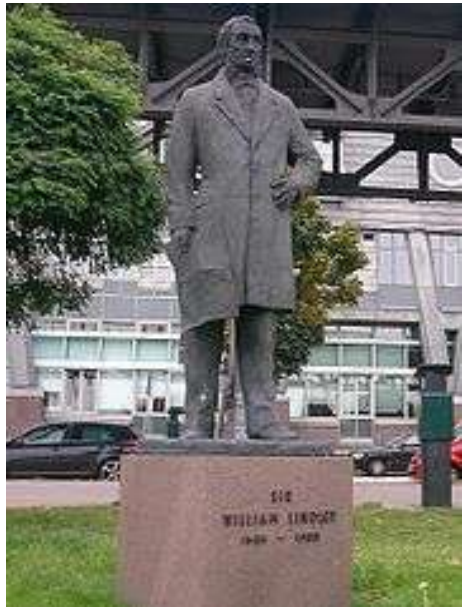

Figure 3: Statue of William Lindley (1808-1900) in Hamburg

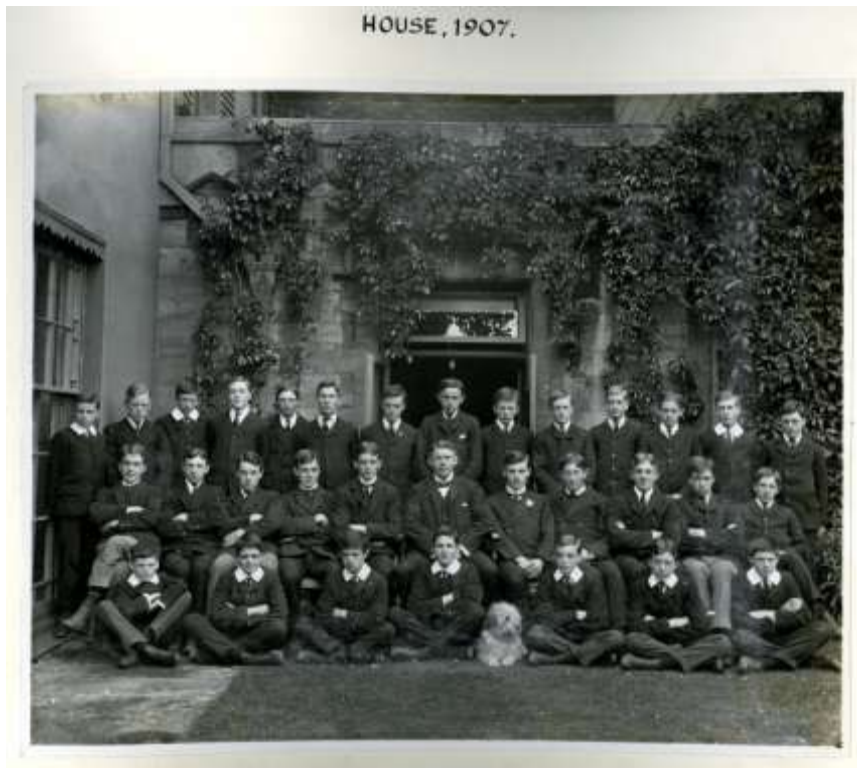

Figure 4: Harper House at Sherborne School in 1907

Lindley is in the front row, second from left (image courtesy of Rachel Hassall, Sherborne School)

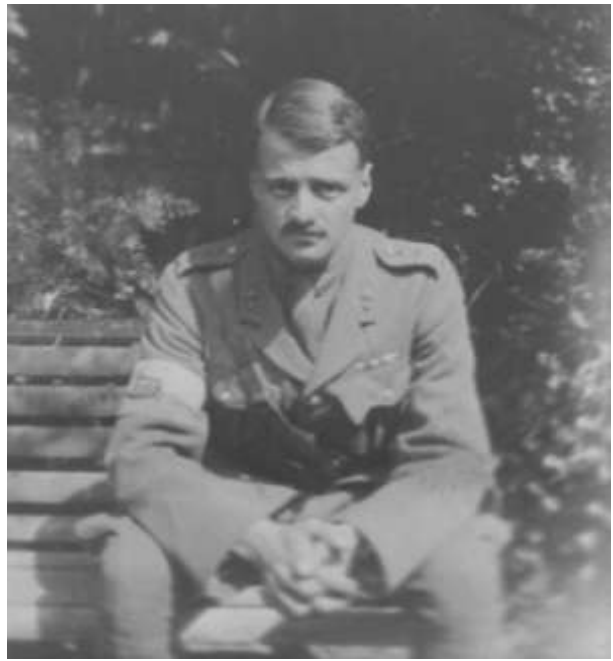

Figure 5: Lindley during World War 1

(image: Morwyn Porter)

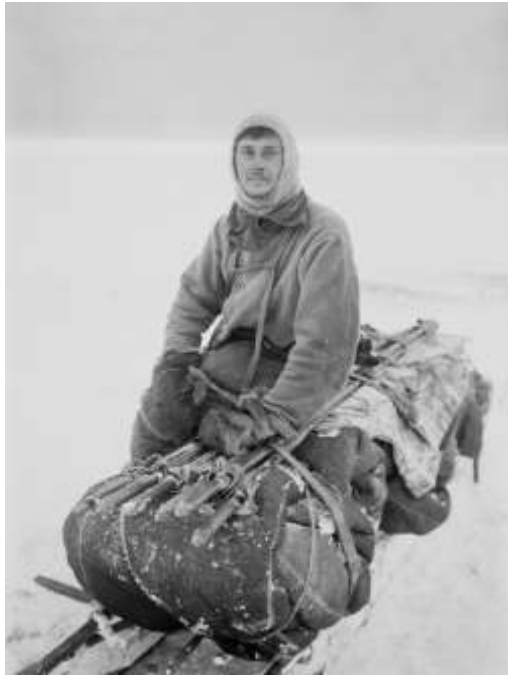

Figure 6: Charles Seymour Wright (1887-1975) on 13 April 1911

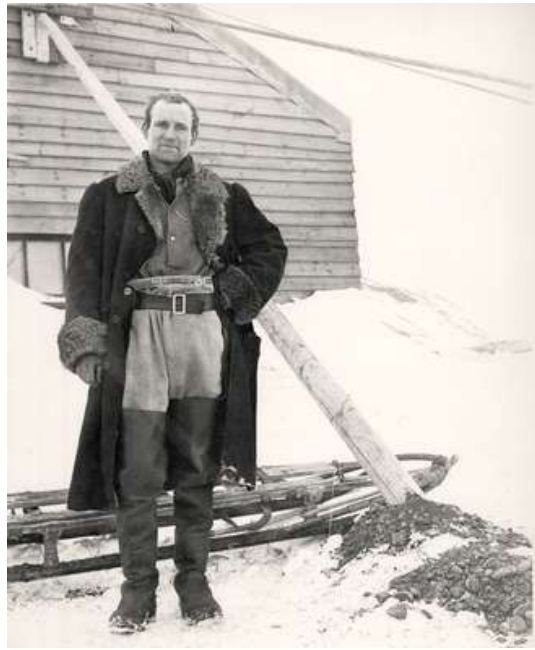

Figure 7: Raymond Priestley (1886-1974)

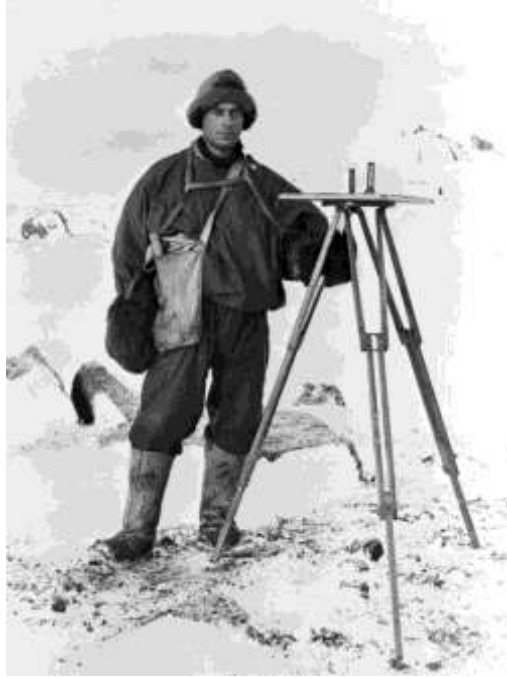

Figure 8: Frank Debenham (1883-1965) on 9 September 1911

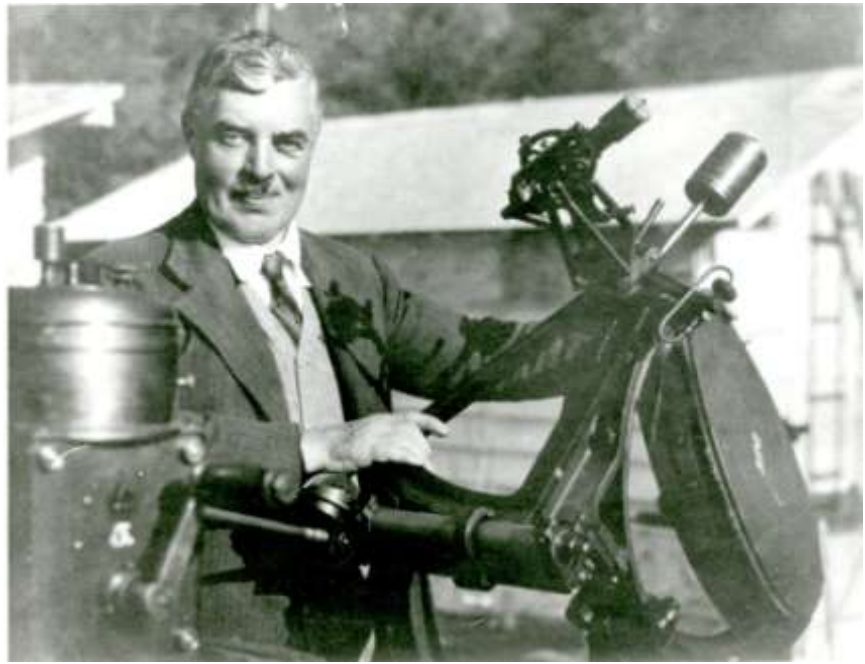

Figure 9: F.J.M. Stratton (1881-1960) at the Japan solar eclipse of June 1936

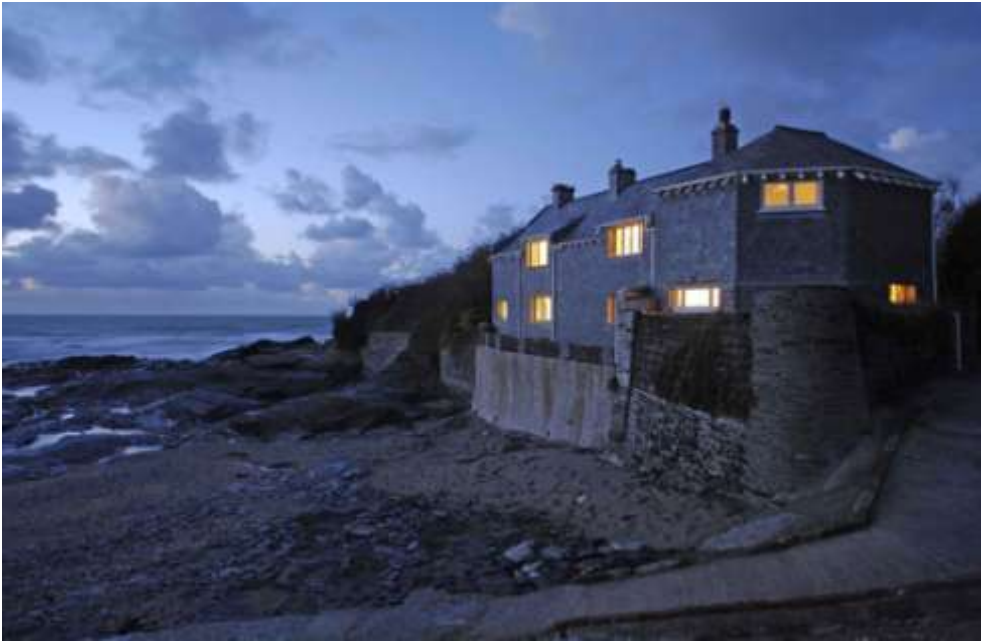

Figure 10: Lindley's residence, Pentonwarra, Trevone, Cornwall

*(image: Unique Home Stays, www.uniquehomestays.com)*

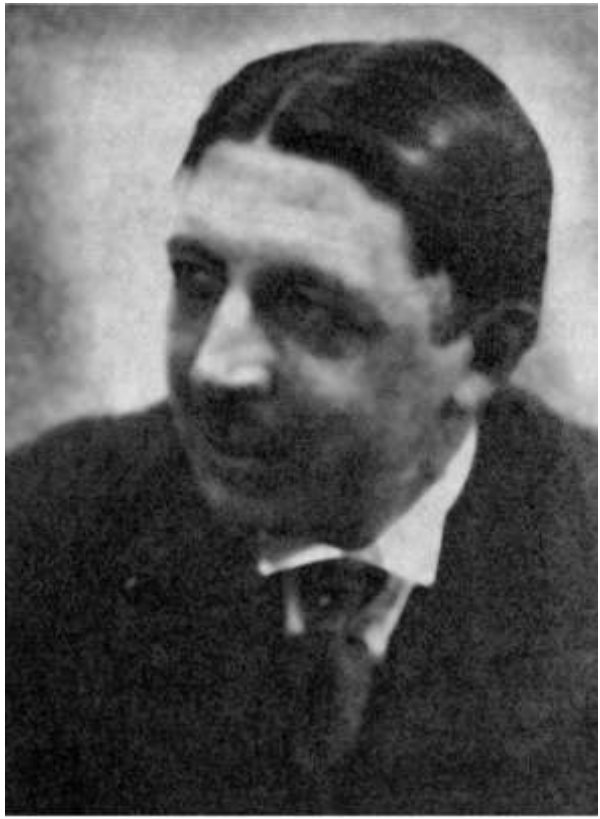

Figure 11: Félix Eugène Marie de Roy (1883-1942)

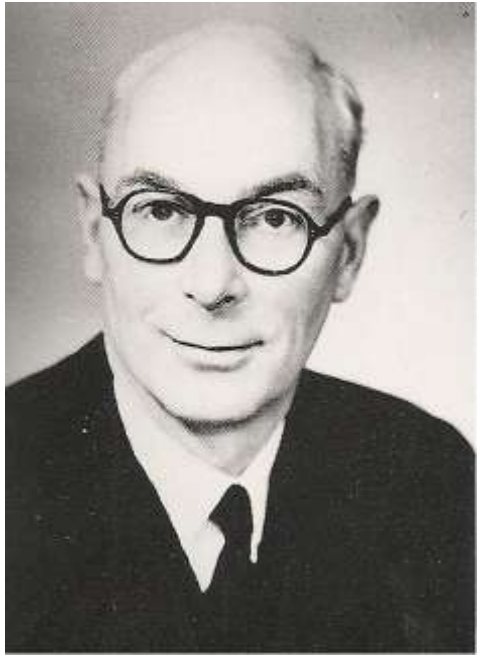

Figure 12: E.H. Collinson (1903-1990)

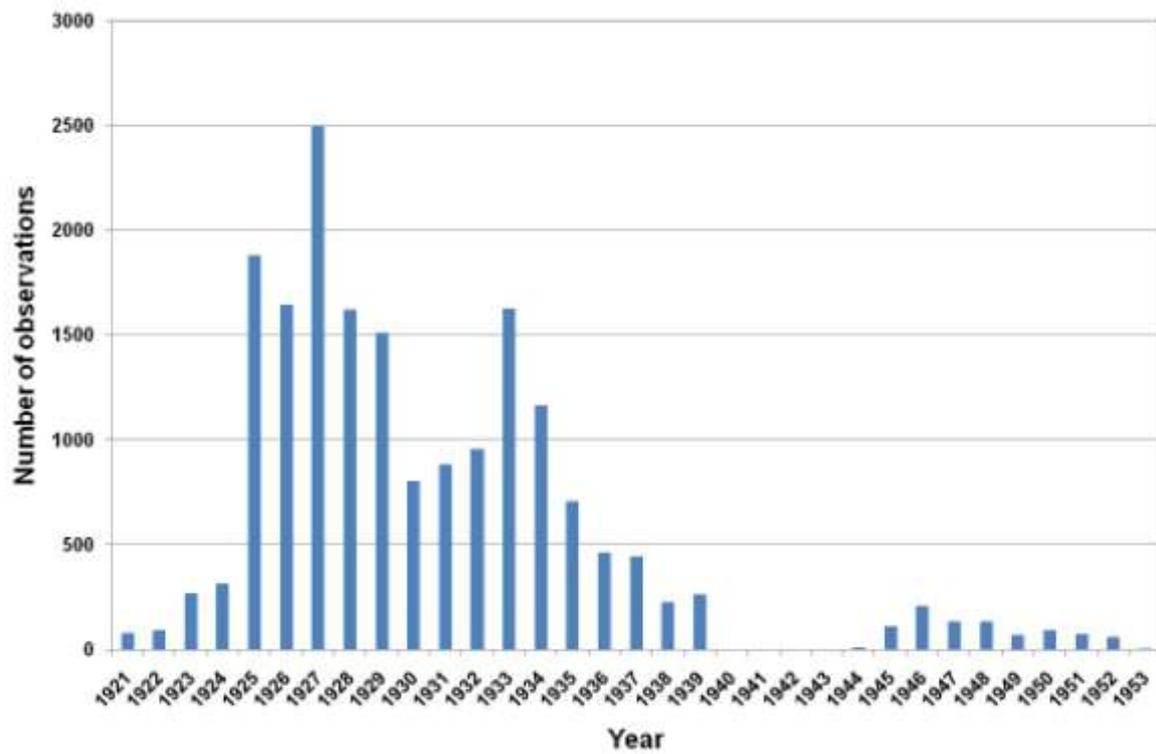

Figure 13: Lindley's variable star observations

*(source: BAA VSS database* (102)*)*

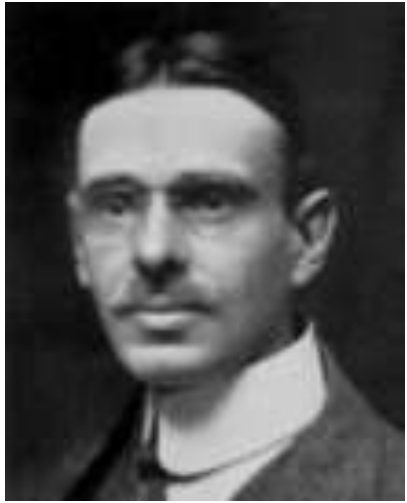

Figure 14: William Tyler Olcott (1873-1936), AAVSO founder

*(AAVSO Archives)*

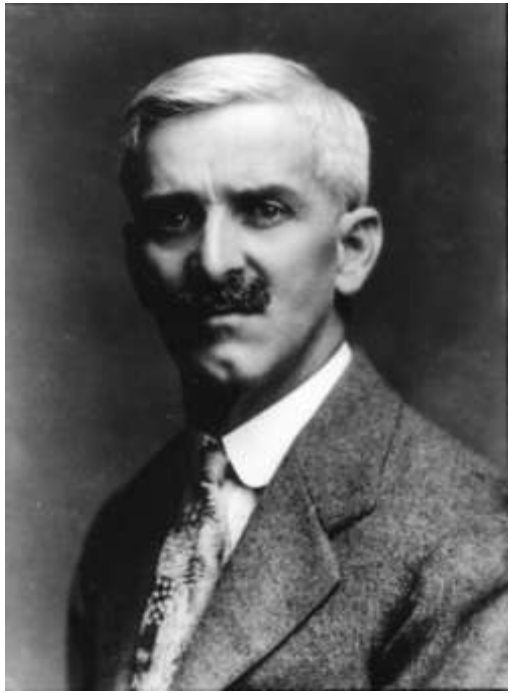

Figure 15: Leon Campbell (1881-1951), Recorder of the AAVSO

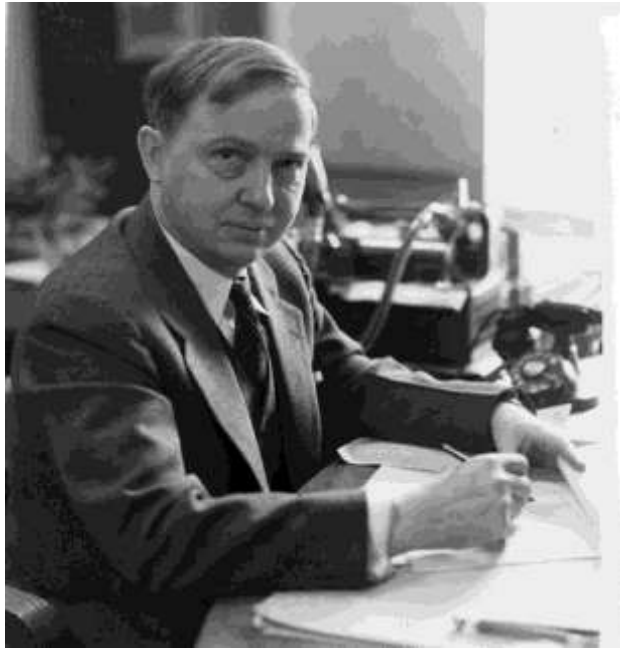

Figure 16: Harlow Shapley (1885-1972), Director of the Harvard College Observatory

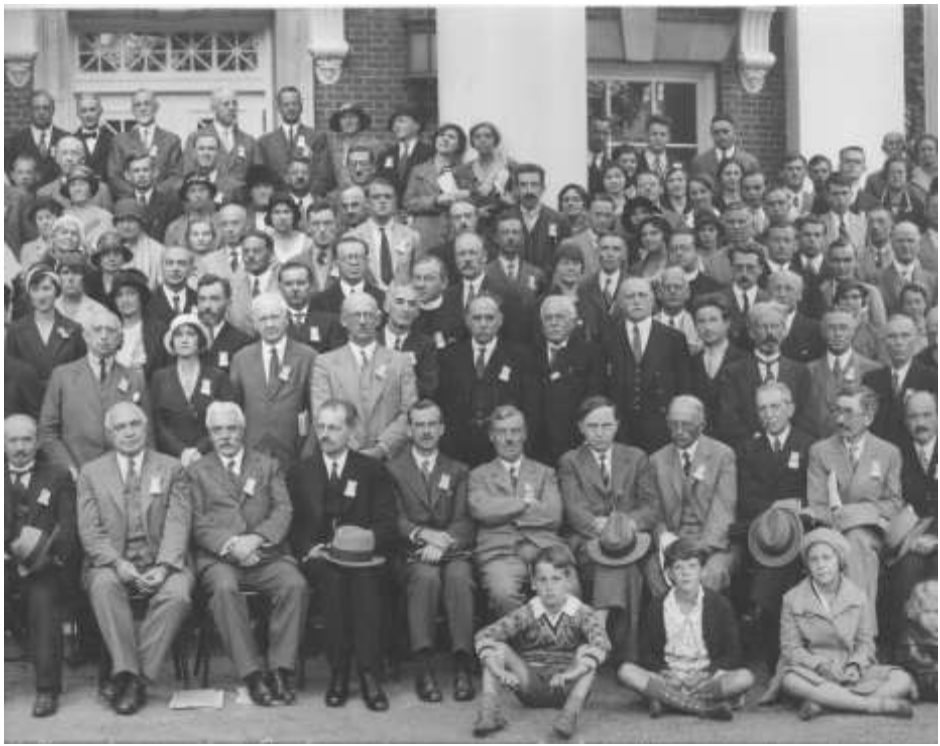

Figure 17: Delegates at the 1932 IAU meeting in Cambridge, Massachusetts

Lindley is in the back row, 5th from left. Mrs Lindley is next to him (6th from left). In the front row, left to right: E. Esclangon (Paris), F. Schlesinger (Yale), F. Dyson (Greenwich), N.E., Nørlund (Copenhagen), G. Abetti (Arcetri), F.J.M. Stratton (Cambridge), H. Shapley (Harvard), H.N. Russell (Princeton), A. de la Baume Pluvinol (Paris), J. Bosler (Marseilles), J. Baillaud (Paris). The astrophysicist G. Lemaître can be seen wearing a dog collar approximately three-quarters of the way from left to right and two-thirds from front to back (103).

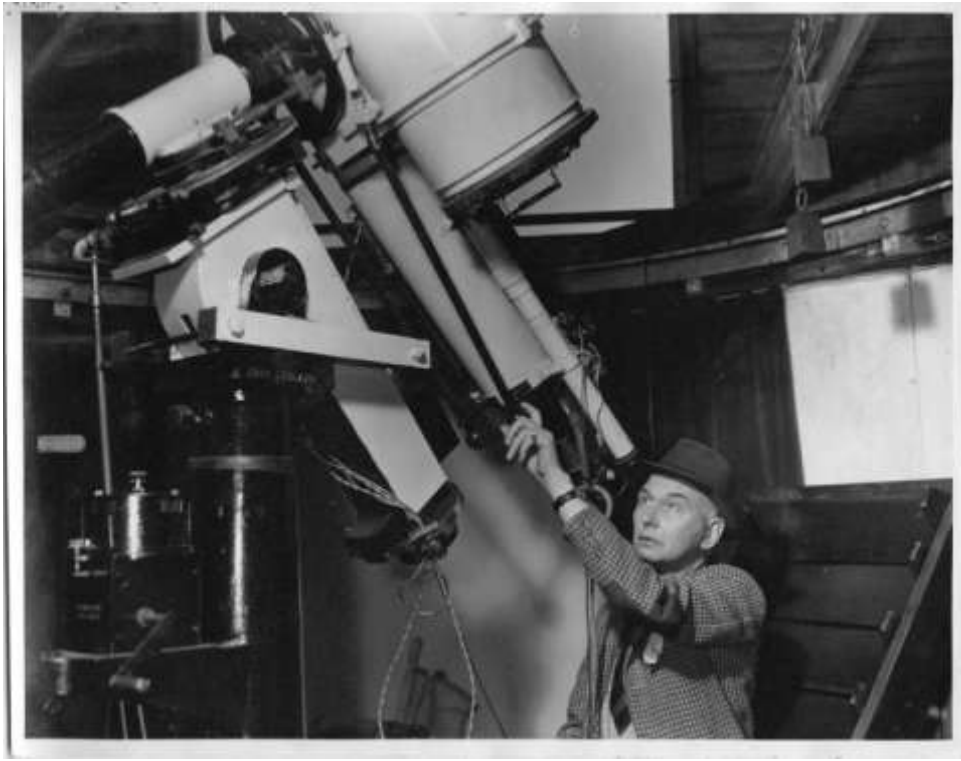

Figure 18: Dr. R.L. Waterfield (1900-1986)

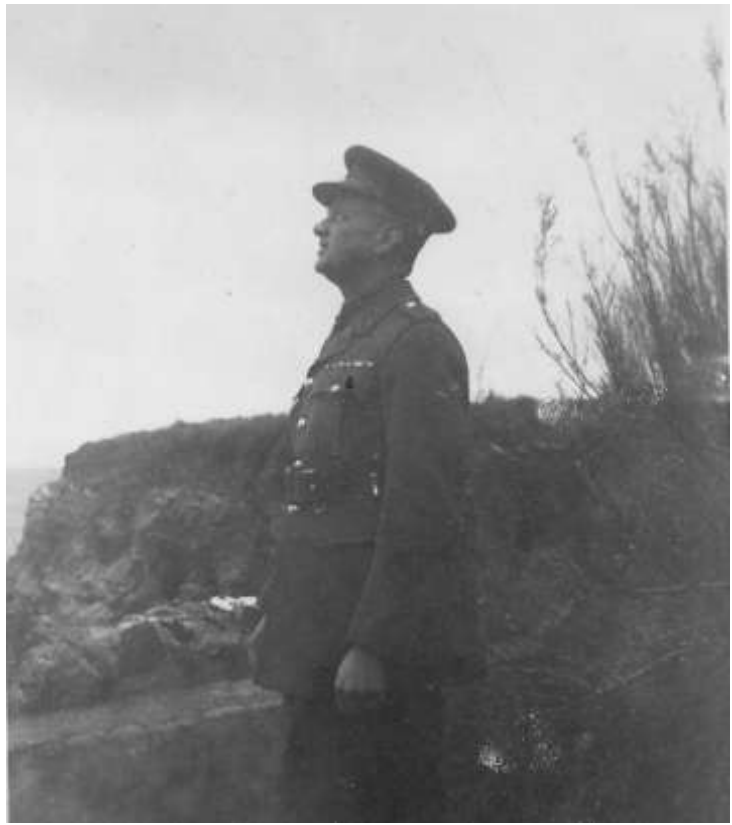

Figure 19: Lindley during World War 2

(image: Morwyn Porter)

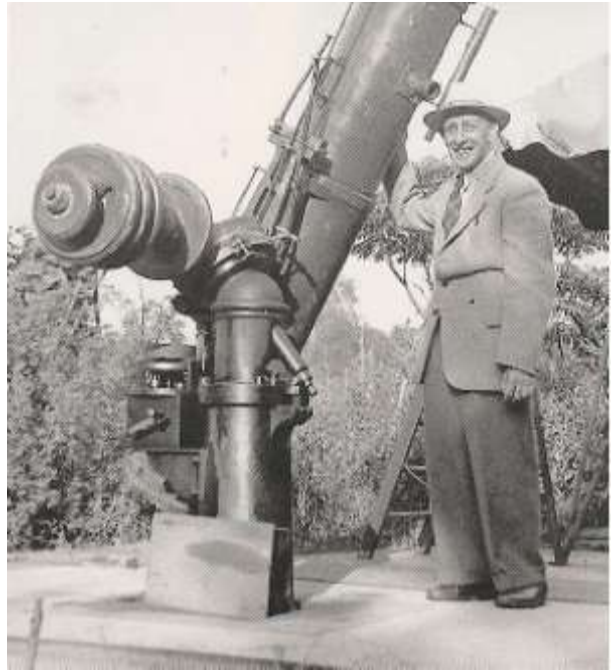

Figure 20: F.M. Holborn (1884-1962)

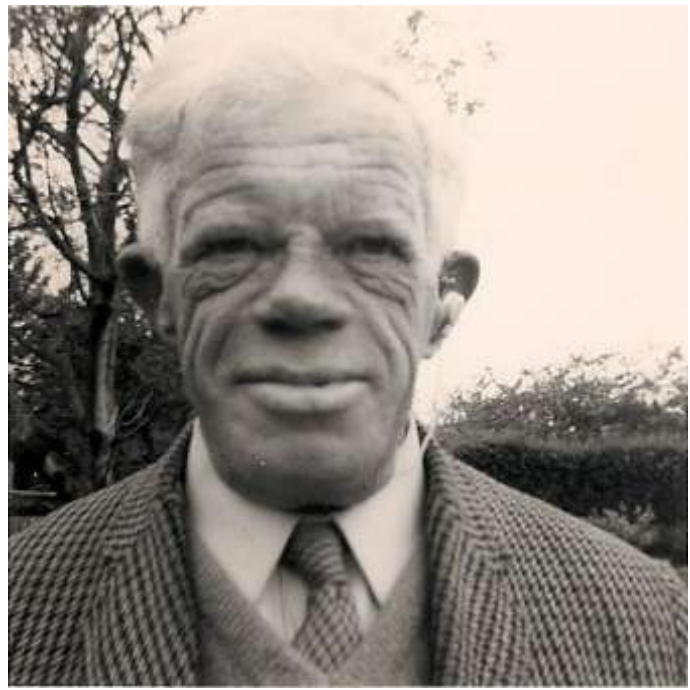

Figure 21: Gordon Patston (1902-1989)

(image: Colin Munford)

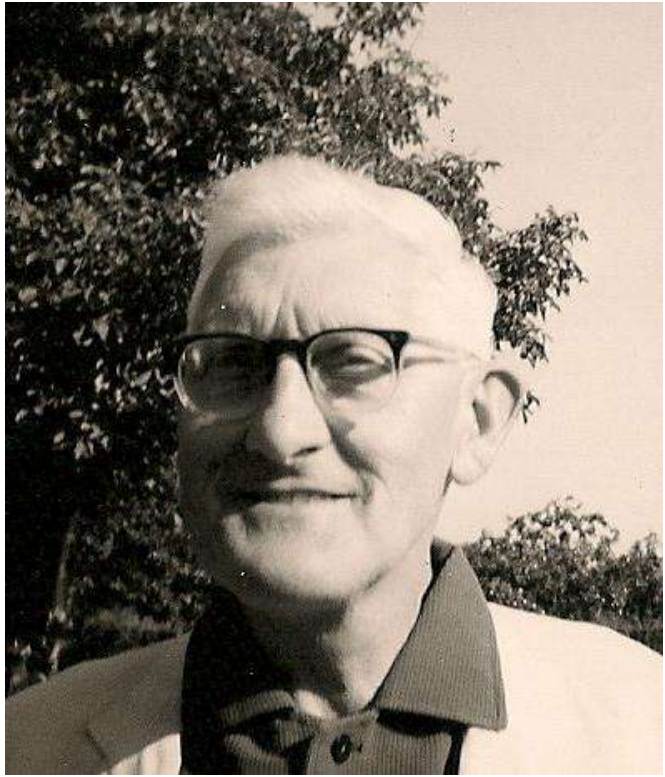

Figure 22: Reginald Gordon Andrews (1903-1996)

(image: Colin Munford)

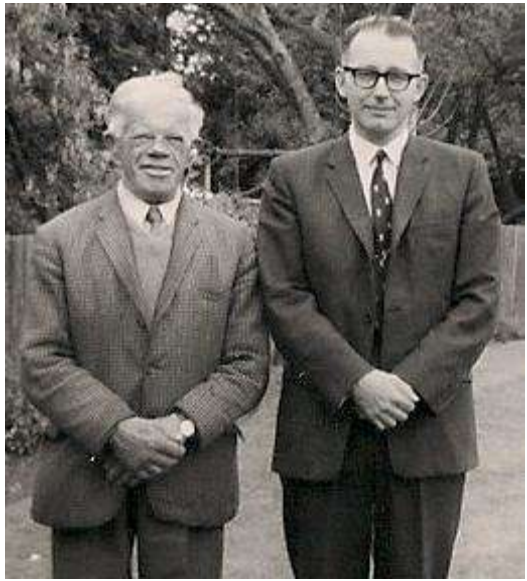

Figure 23: Gordon Patston (left) and Colin Munford (right), 1971

(image: Colin Munford)